\documentclass[11pt]{article}

\usepackage{amsmath,amsfonts,amssymb,bm}

\usepackage{cancel}
\usepackage{tikz-cd}
\usepackage{float}

\usepackage{xcolor}
\usepackage{pifont}%

 \usepackage[numbers,sort&compress]{natbib}
\usepackage{graphicx}

\usepackage{hyperref}
\hypersetup{
   colorlinks   =  true,
    linkcolor    = blue,
    citecolor    = red,
     urlcolor	=magenta,     
}

\usepackage{amsthm}
\theoremstyle{definition} 

\theoremstyle{theorem} 

\newcommand{\sect}[1]{\setcounter{equation}{0}\section{#1}}
\newcommand{\subsect}[1]{\subsection{#1}}
\newcommand{\subsubsect}[1]{\subsubsection{#1}}

\newcommand{\greencheck}{{\color{green}\checkmark}}
\newcommand{\redxmark}{\color{red} \ding{55}}%

\newcommand\be{\begin{equation}}
\newcommand\ee{\end{equation}}
\newcommand\bea{\begin{eqnarray}}
\newcommand\eea{\end{eqnarray}}

\DeclareMathOperator\spn{span}

 \newcommand\st{{\rm st} }
 \newcommand\til{{\rm tl}} 
 \newcommand\sil{{\rm sl}} 
 \newcommand\lil{{\rm ll} }

 \newcommand\timek{{\rm time}}
\newcommand\spa{{\rm space}}
\newcommand\nullplane{{\rm light}}

\newcommand\dd{{\rm d}}

 \def\mf {\mathfrak}

\begin{document}

\

\begin{center}
\baselineskip 24 pt {\LARGE \bf  
All noncommutative spaces  of $\kappa$-Poincar\'e geodesics}
\end{center}

\bigskip
\bigskip

\begin{center}

{\sc Angel Ballesteros$^{1}$, Ivan Gutierrez-Sagredo$^{2}$ and Francisco J.~Herranz$^{1}$}

\medskip
{$^1$Departamento de F\'isica, Universidad de Burgos, 
09001 Burgos, Spain}

{$^2$Departamento de Matem\'aticas y Computaci\'on, Universidad de Burgos, 
09001 Burgos, Spain}
\medskip
 
e-mail: {\href{mailto:angelb@ubu.es}{angelb@ubu.es}, \href{mailto:igsagredo@ubu.es}{igsagredo@ubu.es}, \href{mailto:fjherranz@ubu.es}{fjherranz@ubu.es}}

\end{center}

\medskip

\begin{abstract}

Noncommutative spaces of geodesics provide an alternative way of introducing noncommutative relativistic kinematics endowed with quantum group symmetry. In this paper we  present explicitly the seven noncommutative spaces of 
time-, space- and light-like geodesics that can be constructed from the time-, space- and light- versions of the $\kappa$-Poincar\'e quantum symmetry in (3+1) dimensions. Remarkably enough, only for the light-like (or null-plane) $\kappa$-Poincar\'e deformation the three types of noncommutative spaces of geodesics can be constructed, while for the time-like and space-like deformations both the quantum time-like and space-like geodesics can be defined, but not the light-like one. This obstruction comes from the constraint imposed by the coisotropy condition for the corresponding deformation with respect to the isotropy subalgebra associated to the given space of geodesics, since all these quantum spaces are constructed as quantizations of the corresponding classical coisotropic Poisson homogeneous spaces. The known quantum space of geodesics on the light cone is given by a five-dimensional homogeneous quadratic algebra, and  the six nocommutative spaces of time-like and space-like geodesics are explicitly obtained as six-dimensional nonlinear algebras. Five out of these six spaces are here presented for the first time, and Darboux generators for all of  them are found, thus showing that the quantum deformation parameter $\kappa^{-1}$ plays exactly the same algebraic role on quantum geodesics as the Planck constant $\hbar$ plays in the usual phase space description of quantum mechanics.
\end{abstract}

\medskip
\medskip


\bigskip


\newpage

\tableofcontents

\sect{Introduction}
\label{s1}

It is well-known that quantum groups~\cite{Drinfeld1987icm,ChariPressley1994,majid} provide Hopf algebra deformations of relativistic symmetries that generate noncommutative spacetimes which are covariant under the corresponding quantum kinematical groups of transformations. Then, provided that the quantum deformation parameter is assumed to be related to the Planck scale~\cite{Majid1988}, it is natural to think of quantum Poincar\'e groups as effective (flat spacetime) symmetries in a quantum gravity context, since it is widely assumed that noncommutative spacetimes should arise in the interplay between gravity and quantum effects at such   scale~\cite{Snyder1947,DFR1994,Garay1995,Hossenfelder2013minlength,  Szabo2003,Amelino-Camelia:2008aez,Addazi:2021xuf}. Amongst all possible quantum deformations of the Poincar\'e symmetry, the so-called $\kappa$-Poincar\'e quantum groups and their associated noncommutative $\kappa$-Minkowski spacetimes have focused most of the attention in this direction  (see~\cite{LRNT1991,GKMMK1992,LNR1992fieldtheory,Maslanka1993,MR1994,Zakrzewski1994poincare,CK4d,nullplane95,  BRH2003,Lukierski2006,BorowiecPachol2009jordanian,GM2013relativekappa,BP2014extendedkappa,BGMsymmetry} and references therein). It is also worth stressing that although the original $\kappa$-Poincar\'e deformation is of time-like nature~\cite{LRNT1991,GKMMK1992,LNR1992fieldtheory} (the quantum deformation parameter $\kappa$ is algebraically and thus dimensionally linked to the time translation generator $P_0$), both the space-like~\cite{CK4d} and the light-like (or null-plane)~\cite{nullplane95,Rnullplane97,bicrossnullplane,Tmatrix} $\kappa$-Poincar\'e quantum groups were soon introduced and provided physically different models.

Indeed, classical Minkowski spacetime can be constructed as a homogeneous space of the Poincar\'e group $M=G/H$ where $H$ (the isotropy subgroup) is the Lorentz group ${\rm SO}(3,1)$. From this perspective, $\kappa$-Minkowski spacetimes (time-, space- and light-like ones) can be constructed as the quantization of the Poisson homogeneous space $(M,\pi)$ that arises when $M$ is endowed with the coisotropic Poisson homogeneous structure $\pi$ (see~\cite{LuThesis,Ciccoli2006,BMN2017homogeneous,BGM2019coreductive}) generated by the corresponding classical $r$-matrix (either time-, space- or light-like) that generates the $\kappa$-deformation (see~\cite{BGH2019worldlinesplb} and~\cite{BGH2022light} for a detailed explanation in the time-like and light-like cases, respectively). 

It is also well-known~\cite{HS1997phasespaces,Low1989,BP1991geodesics}, as it will be explained in detail in section \ref{s3}, that the three spaces of geodesics on Minkowski spacetime can also be constructed as homogeneous spaces given as the coset spaces of the Poincar\'e group with respect to the isotropy subgroup for each type of geodesics. Therefore, the quantization of the coisotropic Poisson homogeneous structures induced on them by the three $\kappa$-Poincar\'e $r$-matrices would give rise, in principle, to nine different quantum spaces of geodesics. In this way, the highly nontrivial problem dealing with the definition of the quantum geodesics associated to the $\kappa$-Minkowski noncommutative spacetimes can be solved and, obviously, these novel quantum spaces of geodesics will be, by construction, covariant under the suitable (co)action of the $\kappa$-Poincar\'e quantum group that quantizes the Poisson--Lie Poincar\'e structure under consideration.

However, this fully constructive approach has not been considered in the literature until very recently. In particular, in \cite{BGH2019worldlinesplb} the (six-dimensional (6D)) noncommutative space of time-like geodesics  associated to the time-like $\kappa$-Poincar\'e quantum group was constructed, and some phenomenological implications of this model for the associated `fuzzy' worldlines of free massive particles were further analysed in~\cite{BGGM2021fuzzy}.  Moreover, in \cite{BGH2022light}, the relevant problem concerning the construction of the (5D) space of quantum geodesics on the light cone was fully solved for the light-like (or null-plane) $\kappa$-deformation of the Poincar\'e group. 

Surprisingly enough, in~\cite{BGH2022light} it was also found that both the time-like and the space-like $\kappa$-deformations cannot be used to provide coisotropic Poisson homogeneous structures on the space of geodesics on the light cone, and therefore the two corresponding quantum spaces of light-like geodesics cannot be constructed. This was discovered to be an outstanding property of the space of geodesics on the light cone, since it can also be  proven that the three types of $\kappa$-deformations allow the construction of coisotropic Poisson homogeneous spaces for both the time-like and the space-like spaces of geodesics. Therefore, from the nine possible noncommutative spaces of geodesics coming from the three $\kappa$-Poincar\'e deformations and the three types of geodesics on Minkowski spacetime, two of them are  excluded due to the coisotropy constraint. 

Since only two of the remaining seven spaces have been   deduced so far in~\cite{BGH2019worldlinesplb} and~\cite{BGH2022light}, the aim of this paper is to construct explicitly the five unknown quantum spaces of $\kappa$-Poincar\'e geodesics and to present a global overview of all these seven novel noncommutative structures in a common setting. We think that this comprehensive view will be helpful in order to explore their mathematical interrelations as well as their possible physical features as kinematical models at the Planck scale, that we plan to study in the sequel. 

The structure of the paper is as follows.
In the next two sections, we provide all the essential ingredients of the formalism that will be used in the paper. We start in section \ref{s2} by recalling the  (3+1)D Poincar\'e Lie algebra $\mathfrak g$ expressed in the usual kinematical basis and we introduce the so called null-plane basis~\cite{Leutwyler78}. The $r$-matrices underlying the three classes of  $\kappa$-Poincar\'e deformations (see~\cite{BP2014extendedkappa} and references therein) are given in a covariant way, and their Lie bialgebra structures $(\mathfrak g,\delta)$ are explicitly computed. In section~\ref{s3}, we introduce   the (3+1)D Minkowskian spacetime as well as the  6D time-like, 6D space-like and 5D light-like spaces of geodesics as homogeneous spaces $G/H$ of the Poincar\'e group $G$, where $H$ is the appropriate isotropy subgroup for each space. Then, it is checked for each space whether each of the Lie bialgebra structures associated to the three $\kappa$-deformations fulfils the coisotropy condition $\delta(\mathfrak h) \subset \mathfrak h \wedge \mathfrak g$, where  $\mf h$ is the Lie algebra of the isotropy subgroup $H$. This allows us to identify which  deformations provide a coisotropic Poisson homogeneous structure on a given $G/H$ that will be suitable for quantization. Also, the first-order of all the noncommutative spaces that  satisfy the coisotropy condition are  derived from the associated $\kappa$-Poincar\'e Lie bialgebra structures. 
Finally, the generic approach to the construction of a noncommutative space as the quantization of a coisotropic Poisson homogeneous space is illustrated in detail by recovering through such a procedure the three well-known $\kappa$-Minkowski spacetimes arising from the time-, space- and light-like $\kappa$-deformations. 

This formalism is applied to the construction of all the quantum spaces of geodesics in section \ref{s4} where, in particular, sections  \ref{s41} and \ref{s42} contain the explicit  6D  noncommutative spaces of time- and space-like geodesics, respectively, which arise from the three types of $\kappa$-Poincar\'e quantum groups.  We stress that although the noncommutative space of time-like geodesics induced from the time-like deformation has already been  obtained in~\cite{BGH2019worldlinesplb},     the  other five noncommutative spaces of geodesics here presented are novel ones. It is worth mentioning that all these quantum spaces are no longer of Lie-algebraic type, which means that the first-order structures obtained in section~\ref{s3} by the Lie bialgebra structures  have to be completed in a substantially non-trivial way by including many higher order contributions. Furthermore, Darboux-type coordinates are found for suitable submanifolds in the six coisotropic Poisson homogeneous structures, which implies that for each of these 6D quantum spaces of geodesics there exist appropriate generators for which the defining noncommutative algebra is given by canonical Heisenberg--Weyl relations where the quantum deformation parameter $\kappa^{-1}$ replaces the Planck constant $\hbar$. Indeed, this means that the phenomenological analysis performed in~\cite{BGGM2021fuzzy} could be extended to the other  five  new spaces by following a similar procedure. For the sake of completeness, we sketch  in \ref{s43} the basic features of the 5D noncommutative space of light-like geodesics introduced in~\cite{BGH2022light}, which arises from the light-like $\kappa$-Poincar\'e deformation. Finally, some remarks and open problems are presented in a closing section.


\sect{$\kappa$-Poincar\'e Lie bialgebras}
\label{s2}

Let us consider the  (3+1)D Poincar\'e Lie algebra $\mathfrak{g}=  \mathfrak{iso}(3,1)  \equiv \mathfrak{so}(3,1) \ltimes \mathbb{R}^4$ in  the usual kinematical basis $\{P_0,P_a, K_a, J_a\}$  $(a=1,2,3)$ spanned, in this order,  by the generators of time translation, space translations, boosts and rotations. The   commutation rules for $\mathfrak{g}$ are given by
\be
\begin{array}{llll}
[J_a,J_b]=\epsilon_{abc}J_c ,& \quad [J_a,P_b]=\epsilon_{abc}P_c , &\quad
[J_a,K_b]=\epsilon_{abc}K_c ,  &\quad  [J_a,P_0]=0 , \\[2pt]
\displaystyle{
  [K_a,P_0]=P_a  } , &\quad\displaystyle{[K_a,P_b]=\delta_{ab} P_0} ,    &\quad\displaystyle{[K_a,K_b]=-\epsilon_{abc} J_c} , 
 &\quad 
[P_\mu,P_\nu]=0 ,
\end{array}
\label{aa}
\ee
where   $a,b,c=1,2,3$,  $\mu,\nu=0,1,2,3$, and  the speed of light is set to $c=1$.
Hereafter sum over repeated indices will be assumed, unless otherwise stated.  

In the following it will be useful to consider also the so-called  `null-plane basis' $\{P_\pm,P_i, E_i, F_i,K_3,L_3\}$ $(i=1,2)$~\cite{Leutwyler78},  associated with the boost generator $K_3$ in such a way that the    Poincar\'e  generators   can be casted into three different classes  according to the adjoint action of $K_3$  
\be
[K_3,X]=\gamma X,\qquad X\in \mf g ,
\label{goodness}
\ee
where the parameter $\gamma$ is called  the `goodness' of the generator $X$~\cite{Leutwyler78}. 
Explicitly, the null-plane generators~\cite{BGH2022light} are defined as:
\be
\begin{array}{ll}
\mbox{$\gamma=+1$:}&\quad \displaystyle{ P_+=\frac 12 (P_0+P_3) ,\quad  E_1=\frac 12(K_1-J_2)  ,\quad E_2=\frac 12(K_2 +J_1). }\\[6pt]
\mbox{$\gamma=0$:}&\quad   K_3,\quad L_3=-J_3,\quad P_1,\quad P_2 . \\[5pt]
\mbox{$\gamma=-1$:}&\quad  P_-= 
P_0-P_3,\quad F_1=  K_1
+J_2 ,\quad F_2= K_2 -J_1 \, ,
\end{array}
\label{ad}
\ee
from which the commutation rules of the Poincar\'e algebra are written as:
\be
\begin{array}{llll}
 [L_3,E_i]=-\epsilon_{ij3}E_j,& \quad [L_3,F_i]=-\epsilon_{ij3}F_j , &\quad
[L_3,P_i]=-\epsilon_{ij3}P_j ,  &\quad   [L_3,P_\pm]=0 , \\[2pt]
    [K_3,E_i]=E_i  , &\quad [K_3,F_i]=-F_i,    &\quad [K_3,P_i]=0, 
 &\quad 
[K_3,P_\pm]=\pm P_\pm,\\[2pt]
 [E_i,P_j]=\delta_{ij}P_+, &\quad [E_i,P_+]=0,    &\quad  [E_i,P_-]=P_i, 
 &\quad 
[E_1,E_2]=0,\\[2pt]
 [F_i,P_j]=\delta_{ij}P_-, &\quad [F_i,P_+]= P_i,    &\quad   [F_i,P_-]= 0, 
 &\quad 
[F_1,F_2]=0,\\[2pt]
[K_3,L_3]=0, 
 &\quad 
[P_\alpha,P_\beta]=0,&    \multicolumn{2}{l} {\quad
 [E_i,F_j]=\delta_{ij}K_3 +\epsilon_{ij3}L_3,    }  
 \end{array}
\label{ae}
\ee
where  $i,j=1,2$ and $\alpha, \beta=\pm,1,2$.

We   recall that all   possible quantum  deformations of the Poincar\'e algebra are coboundary ones~\cite{Zakrzewski1995,Zakrzewski1997,PW1996} (except for the very particular (1+1)D case~\cite{Zakrzewski1995,VK,Tmatrix}), which means that all of them  are generated by 
 classical $r$-matrices. The coboundary Lie bialgebra structure 
$(\mf g, \delta)$  associated to a given deformation is determined by the
cocommutator map  $\delta: \mathfrak g\rightarrow \mathfrak g\wedge \mathfrak g$, which is   obtained from a  classical $r$-matrix   $r\in \mathfrak g\wedge \mathfrak g$  in the form
\be
\delta(X) = [X \otimes 1 + 1 \otimes X,r], \quad \forall X \in \mathfrak{g}.
\label{ba}
\ee
Among all   possible $r$-matrices for the Poincar\'e algebra~\cite{Zakrzewski1997}, we shall focus in this paper  on the so-called $\kappa$-Poincar\'e $r$-matrices, where $\kappa^{-1}$ plays the role of the quantum deformation parameter. These $r$-matrices can be written   in covariant notation  as~\cite{Zakrzewski1997, BP2014extendedkappa}
\be
r= v^{\mu}\,  M_{\mu\nu}\wedge P_\nu\, ,
\label{bbaa}
\ee
where $v^{\mu}$ are the components of a Minkowskian four-vector $v=(v^0,v^1,v^2,v^3)$ and $M_{\mu\nu}$ denote a set of generators of the Lorentz group, which are related to  the kinematical basis (\ref{aa}) in the form
\be
M_{0\nu} \equiv K_\nu,\qquad
M_{12}= J_3,
\qquad
M_{23}= J_1,
\qquad
M_{31}= J_2.
\ee
Consequently, according to the time-, space- and light-like nature of the four-vector $v$, three different classes of 
classical $r$-matrices~\cite{BP2014extendedkappa} arise, each of them determining a different type of $\kappa$-Poincar\'e deformation. We display in Table~\ref{table1} a representative of each equivalence class of $r$-matrices determined by a choice of four-vector $v$~\cite{BGH2022light}, which are written  in the covariant notation  (\ref{bbaa}),  in the kinematical basis (\ref{aa}) and in the null-plane one (\ref{ae}).


\begin{table}[t!]
{\small
\caption{\small  \cite{BGH2022light} The three classical $r$-matrices  for the time-, space- and light-like $\kappa$-deformations of the (3+1)D Poincar\'e algebra in  covariant notation   (\ref{bbaa}) as well as in   the kinematical (\ref{aa}) and null-plane (\ref{ae}) bases.}
\label{table1}
  \begin{center}
\noindent 
\begin{tabular}{ r l}
\hline

\hline
\\[-6pt]
\multicolumn{2}{l}{$\bullet$ {Time-like $\kappa$-deformation} from  the time-like vector $v=(1/\kappa,0,0,0)$}\\[4pt]
\quad   $ r_\timek$ &\!\!\!\!\!$= \frac{1}{\kappa} \, M_{0\nu}\wedge P_\nu $  \\[4pt]
  &\!\!\!\!\!$= \frac 1\kappa \left(K_1\wedge P_1 +K_2\wedge P_2 + K_3\wedge P_3 \right) $  \\[4pt]
&\!\!\!\!\!$=  \frac 1\kappa\left(K_3\wedge P_+ +E_1\wedge P_1+ E_2\wedge P_2 \right) -\frac {1}{2\kappa}\left(K_3\wedge P_- -F_1\wedge P_1-F_2\wedge P_2 \right) $  \\[8pt]

\multicolumn{2}{l}{$\bullet$ {Space-like $\kappa$-deformation} from  the   space-like vector $v=(0,0,0,-1/\kappa)$}\\[4pt]
\quad  $ r_\spa$ &\!\!\!\!\!$=   -\frac{1}{\kappa} \, M_{3\nu}\wedge P_\nu $  \\[4pt]
  &\!\!\!\!\!$=  \frac 1\kappa\left(K_3\wedge P_0 +J_1\wedge P_2 -J_2\wedge P_1 \right)$  \\[4pt]
&\!\!\!\!\!$=  \frac 1\kappa\left(K_3\wedge P_+ +E_1\wedge P_1+ E_2\wedge P_2 \right) +\frac {1}{2\kappa}\left(K_3\wedge P_- -F_1\wedge P_1-F_2\wedge P_2 \right)$  \\[8pt]

\multicolumn{2}{l}{$\bullet$ {Light-like $\kappa$-deformation} from  the   light-like vector $v=(1/\kappa,0,0,-1/\kappa)$}\\[4pt]
\quad $ r_\nullplane$ &\!\!\!\!\!$=   \frac{1}{\kappa} (M_{0\nu}\wedge P_\nu - M_{3\nu}\wedge P_\nu) $  \\[4pt]
 &\!\!\!\!\!$=   \frac 1\kappa\left(K_3\wedge P_0 +K_1\wedge P_1 +K_2\wedge P_2 + K_3\wedge P_3 +J_1\wedge P_2 -J_2\wedge P_1 \right)  $  \\[4pt]
&\!\!\!\!\!$= \frac 2\kappa\left(K_3\wedge P_+ +E_1\wedge P_1+ E_2\wedge P_2 \right)$  \\[6pt]
\hline

\hline
\end{tabular}
 \end{center}
}
 \end{table} 



\begin{table}[t!]
{\small
\caption{\small  The three classes of  $\kappa$-Poincar\'e Lie bialgebras  $(\mf g, \delta)$ obtained from the $r$-matrices given  in Table~\ref{table1}, via the relation~(\ref{ba}), in the kinematical basis  (\ref{aa}) for the  time- and space-like deformations, and in the null-plane basis  (\ref{ae})  for the light-like deformation. The index $a=1,2,3$ while $i=1,2$.}
\label{table2}
  \begin{center}
\noindent 
\begin{tabular}{ r l}
\hline

\hline
\\[-6pt]
\multicolumn{2}{l}{$\bullet$ {Time-like $\kappa$-Poincar\'e Lie bialgebra from $ r_\timek$}}\\[4pt]
\quad   $\delta(P_0)$ &\!\!\!\!\!$  = \delta(J_a) = 0  \qquad  \delta(P_a) = \frac{1}{\kappa} P_a \wedge P_0$  \\[4pt]
\quad   $\delta(K_1)$ &\!\!\!\!\!$  = \frac{1}{\kappa} (K_1 \wedge P_0 + J_2 \wedge P_3 - J_3 \wedge P_2)$  \\[4pt]
 \quad   $\delta(K_2)$ &\!\!\!\!\!$  = \frac{1}{\kappa}  (K_2 \wedge P_0 + J_3 \wedge P_1 - J_1 \wedge P_3)$  \\[4pt]
\quad   $\delta(K_3)$ &\!\!\!\!\!$  = \frac{1}{\kappa}  (K_3 \wedge P_0 + J_1 \wedge P_2 - J_2 \wedge P_1) $   \\[8pt]

\multicolumn{2}{l}{$\bullet$ {Space-like $\kappa$-Poincar\'e Lie bialgebra from $ r_\spa$} }\\[4pt]
\quad   $\delta(X)$ &\!\!\!\!\!$  =   0    \qquad X\in\{P_3,K_i,J_3\}  $  \\[4pt]
\quad   $ \delta(Y)$ &\!\!\!\!\!$=\frac 1\kappa\, Y\wedge P_3\qquad Y\in\{P_0,P_i\} $  \\[4pt]
\quad   $\delta(K_3)$ &\!\!\!\!\!$  =\frac 1\kappa\,  (K_3\wedge P_3+K_1\wedge P_1+K_2\wedge P_2)$  \\[4pt]
 \quad   $\delta(J_1)$ &\!\!\!\!\!$  = \frac 1\kappa\, (J_1\wedge P_3- K_2\wedge P_0 -J_3\wedge P_1)$  \\[4pt]
\quad   $\delta(J_2)$ &\!\!\!\!\!$  = \frac 1\kappa\,  (J_2\wedge P_3+K_1\wedge P_0 -J_3\wedge P_2) $   
 \\[8pt]

\multicolumn{2}{l}{$\bullet$ {Light-like $\kappa$-Poincar\'e Lie bialgebra from $ r_\nullplane$}}\\[4pt]
\quad   $\delta(X)$ &\!\!\!\!\!$  =   0    \qquad X\in\{P_+,E_i,L_3\}$  \\[4pt]
\quad   $\delta(Y)$ &\!\!\!\!\!$  =\frac 2\kappa\, Y\wedge P_+\qquad Y\in\{P_-,P_i\}$  \\[4pt]
\quad   $\delta(K_3)$ &\!\!\!\!\!$  =\frac 2\kappa\,  (K_3\wedge P_++E_1\wedge P_1+E_2\wedge P_2)$  \\[4pt]
 \quad   $\delta(F_1)$ &\!\!\!\!\!$  = \frac 2\kappa\, (F_1\wedge P_++E_1\wedge P_- +L_3\wedge P_2)$  \\[4pt]
\quad   $\delta(F_2)$ &\!\!\!\!\!$  = \frac 2\kappa\,  (F_2\wedge P_++E_2\wedge P_- -L_3\wedge P_1) $   \\[6pt]
\hline

\hline
\end{tabular}
 \end{center}
}
 \end{table} 


From the explicit expressions presented in Table~\ref{table1},  it becomes clear that both time- and space-like $\kappa$-deformations are naturally adapted to the kinematical basis  (\ref{aa}), while the   light-like $\kappa$-deformation can be better understood in the null-plane basis (\ref{ae}). Moreover, let us note that, as expected
\be
  r_\nullplane=r_\timek+r_\spa .
  \label{bbaa2}
  \ee
The corresponding cocomutator map $\delta$ coming from (\ref{ba}) is explicitly  written in Table~\ref{table2} in the most adapted basis for each deformation, thus providing the three classes of  $\kappa$-Poincar\'e Lie bialgebras $(\mf g, \delta)$. As it is well-known, each of them provides the first-order deformation in $\kappa^{-1}$ of the coproduct of the associated quantum Poincar\'e algebra.

From either Table~\ref{table1} or Table~\ref{table2}, it comes out that the time-, space- and light-like deformation is characterized   by the primitive ({\em i.e.}, with nondeformed coproduct) generator $P_0$, $P_3$ and $P_+$, respectively.  In particular,  $r_\timek$ corresponds to the well-known $\kappa$-Poincar\'e algebra~\cite{LRNT1991,GKMMK1992,LNR1992fieldtheory,Maslanka1993,MR1994,Zakrzewski1994poincare} for which  the deformation parameter $\kappa$ has  dimensions of ${\rm time}^{-1}$ (recall that $c=1$).  The second Lie bialgebra, determined by $r_\spa$,  underlies  the   $q$-Poincar\'e algebra obtained in~\cite{CK4d} (c.f.~Type 1.~(a) with  $z=1/\kappa$), with    $\kappa$   having dimensions of ${\rm length}^{-1}$. Both $r_\timek$ and $r_\spa$ lead to quasitriangular (or standard) deformations of the Poincar\'e algebra since they are solutions of the modified classical Yang--Baxter equation (with non-vanishing Schouten bracket).  The third Lie bialgebra structure, coming from   $r_\nullplane$,  provides    the null-plane quantum Poincar\'e algebra
 introduced in~\cite{nullplane95,Rnullplane97,bicrossnullplane} (where $z=1/\kappa$), which    is a triangular (or nonstandard) quantum deformation 
 with vanishing  Schouten bracket. Therefore, despite its apparent formal similarity, the third $r$-matrix  $r_\nullplane$ (and therefore its Lie bialgebra) is completely different from the other two from a kinematical viewpoint, and this fact will be essential as far as the construction of the corresponding noncommutative spaces of geodesics is concerned.


\newpage

\sect{Homogeneous spaces, the coisotropy condition and quantization}
\label{s3}

The (3+1)D Poincar\'e group $G={\rm ISO}(3,1)$ with Lie algebra  $\mathfrak{g}=  \mathfrak{iso}(3,1) $ allows the construction of several 
 $\ell$D homogeneous spaces which can be  expressed in a generic form as  a    left coset space
\be
M^\ell=G/{H}
\label{tp1}
\ee 
 where $M^\ell$ is the  $\ell$D manifold and   $H$ is the $(10-\ell)$D isotropy subgroup  with   Lie algebra  $\mathfrak{h}$. 
   We can identify the tangent space at every point $m = g H \in M^\ell$,  $g\in G$,  with the translation sector:
\begin{equation}
T_{m} (M^\ell) = T_{gH} (G/H) \simeq \mathfrak{g}/\mathfrak{h} \simeq \mathfrak{t} = \spn{ \{T_1,  \dots, T_\ell\}} .
\label{tp22}
\end{equation}
In fact,  at a Lie algebra level, such construction comes from the Cartan decomposition of the 
   Poincar\'e   algebra $\mathfrak{g}$, as a vector space,   given by the   sum of two subspaces
\be
{\mathfrak{g}}=  \mathfrak{h} \oplus \mathfrak{t} , \qquad  [\mathfrak{h} ,\mathfrak {h} ] \subset \mathfrak{h}  .
\label{tp2}
\ee
Hence the generators of the isotropy subalgebra $\mathfrak{h}$ leave a point on $M^\ell$ invariant, which is taken as the origin $O$  of the homogenous space, thus   playing the role of `rotations' around $O$,  while the $\ell$ generators belonging to $\mathfrak{t}=\spn\{ T_1,\dots, T_\ell\}$ move $O$  along  $\ell$ basic directions,  thus behaving as translations on   $M^\ell$. When appropriately defined, the   local coordinates $(t^1,\dots,t^\ell)$  of the Poincar\'e group   associated with the translation generators of $\mathfrak{t}$ descend to  $\ell$ coordinates  on $M^\ell$.

We    consider in this paper  the four  most relevant homogeneous spaces coming from the Poincar\'e group, namely the Minkowski spacetime and the three types of spaces of geodesics, which are explicitly defined as follows:
\be
\begin{array}{lll}
\multicolumn{3}{l}{ \!\!\!\!\! \mbox{$\bullet$ The (3+1)D Minkowski spacetime} \ \mathcal{M}  =G/H_\st\!:}\\[3pt]
\mathfrak g= \mathfrak t_\st \oplus \mathfrak h_\st ,  \quad & \mathfrak{t}_\st = \spn \{P_0,  {P_a} \}    ,\  & \mathfrak h_\st = \spn\{ {K_a},  {J_a} \}=\mathfrak{so}(3,1) .\\[8pt]
\multicolumn{3}{l}{ \!\!\!\!\! \mbox{$\bullet$ The 6D space of time-like lines} \ \mathcal{W}_\til  =G/H_\til\!:}\\[3pt]
\mathfrak g= \mathfrak t_\til \oplus \mathfrak h_\til ,  \quad & \mathfrak{t}_\til = \spn \{  {P_a},  {K_a} \}    ,\   & \mathfrak h_\til = \spn\{P_0,  {J_a} \}=\mathbb R\oplus \mathfrak{so}(3) .\\[8pt]
\multicolumn{3}{l}{ \!\!\!\!\! \mbox{$\bullet$ The 6D space of space-like lines} \  \mathcal{W}_\sil  =G/H_\sil\!:}\\[3pt]
\mathfrak g= \mathfrak t_\sil \oplus \mathfrak h_\sil ,  \quad & \mathfrak{t}_\sil = \spn \{P_0,P_i,  K_3,J_i\}    ,\   & \mathfrak h_\sil = \spn\{P_3,  K_i,J_3 \}=\mathbb R\oplus \mathfrak{so}(2,1) .\\[8pt]
\multicolumn{3}{l}{ \!\!\!\!\! \mbox{$\bullet$ The 5D space of light-like lines} \    \mathcal{L}  =G/H_\lil\!:}\\[3pt]
\mathfrak g= \mathfrak t_\lil \oplus \mathfrak h_\lil ,  \quad & \mathfrak{t}_\lil = \spn \{P_-,P_i,F_i\}    ,\   & \mathfrak h_\lil = \spn\{P_+,E_i,K_3,L_3 \}  ,
\end{array}
\label{ai} 
\ee
where $a=1,2,3$,  $i=1,2$ and the notation  `$\st$', `$\til$', `$\sil$' and `$\lil$' means, in this order, spacetime, time-like, space-like and light-like. 

\subsect{The coisotropy condition}
\label{s31}

The method that we propose in order to construct quantum group invariant noncommutative analogues of the aforementioned homogeneous spaces $M^\ell$ consists in quantizing the unique coisotropic Poisson homogeneous structure $\pi$ on $M^\ell$ that is covariant under the Poisson--Lie Poincar\'e group that is defined by the $r$-matrix underlying the chosen quantum deformation. The quantization of such Poisson-noncommutative structure on the classical space $M^\ell$ will then provide the defining relations for the quantum space that is covariant under the given quantum deformation. However, as it was discussed in detail in~\cite{LuThesis,Ciccoli2006,BMN2017homogeneous,BGM2019coreductive,GH2021symmetry}, the construction of such Poisson structure $\pi$ is only   guaranteed   whenever the so-called coisotropy condition for the cocommutator $\delta$ with respect to the isotropy subalgebra $\mathfrak h$ of $H$ holds, namely
 \be
\delta(\mathfrak h) \subset \mathfrak h \wedge \mathfrak g.
\label{coisotropy}
\ee
In the particular case when
\be
\delta\left(\mathfrak{h}\right) \subset \mathfrak{h} \wedge \mathfrak{h} ,
\label{coisotropy2}
\ee
the condition is obviously fulfilled, but in such a way that the isotropy subalgebra $\mathfrak h $ is also a  sub-Lie bialgebra $(\mathfrak{h},\delta |_\mathfrak{h})$ of $(\mathfrak g,\delta)$. This means that, after quantization, the isotropy subgroup $H$ will be promoted to a quantum subgroup. 

The verification (or not) of the required  coisotropy condition~(\ref{coisotropy}) for the three classes of $\kappa$-Poincar\'e Lie bialgebras and for all the homogeneous spaces described in~\eqref{ai} can be straightforwardly obtained from the explicit expressions of the cocommutator map $\delta$ given in Table~\ref{table2} and from the definition of the corresponding isotropy subalgebras $\mathfrak h$. 

These results are summarized in Table~\ref{table3}   showing that the three classes of $\kappa$-Poincar\'e algebras can be used to provide a noncommutative $\kappa$-Minkowskian spacetime since the coisotropy condition is always fulfilled. However,  we stress that the only quantum $\kappa$-Poincar\'e algebra that enables the construction of the noncommutative counterpart of the  {four}  homogeneous spaces (\ref{ai})  is just the light-like (or null-plane) quantum Poincar\'e algebra, since both the (usual)  time-like and  the space-like $\kappa$-Poincar\'e deformations do not  always satisfy the coisotropy condition, which precludes the construction of their associated light-like quantum spaces of geodesics.


\begin{table}[t]
{\small
\caption{\small \cite{BGH2022light} Coisotropy condition (\ref{coisotropy}) for the three   $\kappa$-Poincar\'e Lie bialgebras given in Table~\ref{table2} with respect to the four different  isotropy subalgebras (\ref{ai}) that ensures   the existence  \greencheck (or not {\redxmark}) of  a  noncommutative Minkowskian spacetime (st) and a  noncommutative  space of  time-like (tl), space-like (sl) and light-like (ll)   geodesics.}
\label{table3}
  \begin{center}
\noindent 
\begin{tabular}{ l l l l l}
\hline

\hline
\\[-8pt]
\qquad    &$\mathfrak{h}_\st $ & $\mf h_\til$ & $\mf h_\sil$& $\mf h_\lil$ \\[4pt]
\hline
\\[-8pt] 
$r_\timek$\qquad\  & $\delta(\mf h_\st)\subset \mf h_\st\wedge \mf g$ \greencheck \qquad\   & $\delta(\mf h_\til)=0$ \greencheck \qquad\  & $\delta(\mf h_\sil)\subset \mf h_\sil\wedge \mf g $ \greencheck \qquad\  & $\delta(\mf h_\lil)\not\subset \mf h_\lil\wedge \mf g$ \redxmark \\[4pt]
$r_\spa$ \qquad\  & $\delta(\mf h_\st)\subset \mf h_\st\wedge \mf g$ \greencheck \qquad\  &  $\delta(\mf h_\til)\subset \mf h_\til\wedge \mf g$ \greencheck \qquad\  &  $\delta(\mf h_\sil)=0$ \greencheck \qquad\  & $\delta(\mf h_\lil)\not\subset \mf h_\lil\wedge \mf g$ \redxmark \\[4pt]
$r_\nullplane $ \qquad\ &$\delta(\mf h_\st)\subset \mf h_\st\wedge \mf g$ \greencheck\qquad\   & $\delta(\mf h_\til)\subset \mf h_\til\wedge \mf g$ \greencheck\qquad \ & $\delta(\mf h_\sil)\subset \mf h_\sil\wedge \mf g$ \greencheck \qquad\  & $\delta(\mf h_\lil)\subset \mf h_\lil\wedge \mf g$ \greencheck   \\[4pt]
\hline

\hline
\end{tabular}
 \end{center}
}
 \end{table}



\subsect{From coisotropic Lie bialgebras to first-order noncommutative spaces}
\label{s32}

It is worth stressing that the first-order in the local coordinates of the noncommutative spaces $M^\ell_\kappa$ can be  deduced directly from the cocommutators written in Table~\ref{table2} by means of  the   dual map $\delta^\ast:\mathfrak{g}^\ast\otimes \mathfrak{g}^\ast \to \mathfrak{g}^\ast$, which is a Lie bracket on the dual Poincar\'e algebra $\mathfrak{g}^\ast$ of $\mathfrak{g}$. 

In our case, let us denote the   quantum or noncommutative  translation coordinates by $(\hat t^1,\dots, \hat t^\ell)\in \mathfrak g^\ast$   corresponding to the classical  Poincar\'e group local coordinates $(t^1,\dots,t^\ell)$ of the translation sector 
$\mathfrak{t}=\spn\{ T_1,\dots, T_\ell\}$. The duality between the generators of $\mathfrak{t}$ and  the quantum coordinates  $(\hat t^1,\dots, \hat t^\ell)$  is determined by a canonical pairing given by the bilinear form
\be
\langle  \hat t^j,T_k \rangle=\delta_k^j  \,,\qquad \forall j,k.
\label{tdd}
\ee
 Let us consider a   given cocommutator $\delta$ in Table~\ref{table2}.  If the coisotropy condition (\ref{coisotropy})  is satisfied, then  the quantum coordinates   $(\hat t^1,\dots, \hat t^\ell)$  close a Lie algebra which is just the   annihilator $\mathfrak{h}^\perp$ of $\mathfrak{h}$ on the dual Poincar\'e algebra $\mathfrak g^*$, and  therefore this defines   
the  noncommutative space 
\be
\mathfrak{h}^\perp \equiv M^\ell_\kappa
\label{tdd2}
\ee
at the first-order in such quantum coordinates (see~\cite{BGM2019coreductive} for details). In fact, it is worth stressing that this is just the meaning of the coisotropy condition in algebraic terms: when~\eqref{coisotropy} is not fulfilled, the first-order commutation rules given by dualizing the Lie bialgebra cocommutator are such that the generators of the space under consideration do not close on a subalgebra since they include other dual generators that correspond to transformations which do not correspond to translations on the chosen space. 

In particular, if we denote by $(\hat x^\mu, \hat \xi^a, \hat \theta^a)$ $(\mu=0,1,2,3;\ a=1,2,3)$ the quantum coordinates dual to the kinematical generators $(P_\mu, K_a, J_a)$  (\ref{aa}),  and 
 by  $(\hat x^\alpha, \hat e^i, \hat f^i, \hat \xi^3, \hat \phi^3)$  $(\alpha=\pm,1,2;\ i=1,2)$ those dual to the   generators in the null-plane basis $(P_\alpha, E_i, F_i,K_3,L_3)$  (\ref{ae}), then the corresponding  first-order quantum spaces (\ref{tdd2})  corresponding to the classical homogenous spaces  (\ref{ai}) will be: 
 \be
\begin{array}{lll}
\multicolumn{3}{l}{ \!\!\!\!\! \mbox{$\bullet$ The (3+1)D $\kappa$-Minkowski spacetime} \ \mathcal{M}_\kappa= \mathfrak h_\st^\perp  = \spn\{\hat x^\mu\}. }  \\[4pt]
\multicolumn{3}{l}{ \!\!\!\!\! \mbox{$\bullet$ The 6D $\kappa$-space of time-like lines}\  \mathcal{W}_{\til,\kappa}= \mathfrak h_\til^\perp  = \spn\{\hat x^a,\hat \xi^a\}. }\\[4pt]
\multicolumn{3}{l}{ \!\!\!\!\! \mbox{$\bullet$ The 6D $\kappa$-space of space-like lines}\  \mathcal{W}_{\sil,\kappa}= \mathfrak h_\sil^\perp  = \spn\{\hat x^0,\hat x^i, \hat \xi^3,\hat \theta^i\}. } \\[4pt]
\multicolumn{3}{l}{ \!\!\!\!\! \mbox{$\bullet$ The 5D  $\kappa$-space of light-like lines}\  \mathcal{L}_{\kappa}= \mathfrak h_\lil^\perp  = \spn\{\hat x^-,\hat x^i, \hat f^i\}. } \\[4pt]
\end{array}
\label{ai2} 
\ee
The explicit expressions of these four first-order noncommutative spaces are displayed in Table~\ref{table4} for each $\kappa$-deformation.
The three noncommutative Minkowski spacetimes $\mathcal{M}_\kappa$ appear in the first line and turn out to have no higher order contributions when the full quantum space is computed, as we will see in the next section (see~\cite{Maslanka1993} for $r_\timek$, \cite{GH2021symmetry} for $r_\spa$,  and  \cite{Rnullplane97}   for  $r_\nullplane$, as well as~\cite{BP2014extendedkappa}). Thus,   they are noncommutative spacetimes of Lie-algebraic type and, furthermore, all of them are isomorphic as Lie algebras (although their physical interpretation is different).

As we will see in section \ref{s4}, this will be no longer the case for the quantum spaces of geodesics, which will be  defined in all the cases by nonlinear relations in terms of the local coordinates on the appropriate parametrization of the Poincar\'e group.  We recall that among the {seven} possible complete  noncommutative spaces of geodesics that are allowed by the coisotropy condition, only two of them have been constructed so far, namely
     $\mathcal{W}_{\til,\kappa}$ from  $r_\timek$ in~\cite{BGH2019worldlinesplb} and   $\mathcal{L}_{\kappa}$  from  $r_\nullplane$ in~\cite{BGH2022light}. Surprisingly enough, we realize that although such two noncommutative spaces are in fact commutative ones at  the  first-order in the quantum coordinates (they are Abelian algebras in Table~\ref{table4}), we will see that they will be defined as noncommutative algebras when contributions at all orders in the quantum coordinates are considered. Namely,
the full quantum space $\mathcal{W}_{\til,\kappa}$ from  $r_\timek$ involves cumbersome expressions with hyperbolic trigonometric functions, meanwhile the complete $\mathcal{L}_{\kappa}$  from  $r_\nullplane$  is defined by quadratic relations.


\begin{table}[t]
{\small
\caption{\small  Non-vanishing commutation relations that define the first-order   noncommutative spaces  $\mathfrak{h}^\perp \equiv M^\ell_\kappa$ corresponding to the four Poincar\'e homogeneous spaces (\ref{ai}) obtained  from the three classes of  $\kappa$-Poincar\'e Lie bialgebras  $(\mf g, \delta)$ given in Table~\ref{table2} in agreement with the coisotropy condition shown in Table~\ref{table3}. The index $a=1,2,3$ while $i=1,2$.}
\label{table4}
  \begin{center}
\noindent 
\begin{tabular}{ l l l l}
\hline

\hline
\\[-8pt]
Space & $r_\timek$ &$r_\spa$  & $r_\nullplane$  \\[4pt]
\hline
\\[-8pt]
$\mathcal{M}_\kappa$ & $  [\hat x^{a},\hat x^{0}] = \frac {1}{\kappa}  \hat x^{a} $  & $  [\hat x^{0},\hat x^{3}] = \frac {1}{\kappa}  \hat x^{0}\quad\   [\hat x^{i},\hat x^{3}] = \frac {1}{\kappa}  \hat x^{i} \   \ $ & $ [\hat x^{-},\hat x^{+}] = \frac {2}{\kappa}  \hat x^{-}    \quad \  [\hat x^{i},\hat x^{+}] = \frac {2}{\kappa}  \hat x^{i}   $ \\[4pt]
$\mathcal{W}_{\til,\kappa}$ &   0&  $  [\hat x^{i},\hat x^{3}] = \frac {1}{\kappa}  \hat x^{i}\quad\   [\hat \xi^{a},\hat x^{a}] = \frac {1}{\kappa}  \hat \xi^{3} \   $ & $  [\hat x^{i},\hat x^{3}] = \frac {1}{\kappa}  \hat x^{i}\quad\  [\hat \xi^{a},\hat x^{a}] = \frac {1}{\kappa}  \hat \xi^{3}    $\\[4pt]
$\mathcal{W}_{\sil,\kappa}$ &  $  [\hat x^{i},\hat x^{0}] = \frac {1}{\kappa}  \hat x^{i}\quad\    [\hat \xi^{3},\hat x^{0}] = \frac {1}{\kappa}  \hat \xi^{3} \    $ & 0 &  $  [\hat x^{i},\hat x^{0}] = \frac {1}{\kappa}  \hat x^{i}\quad \,   [\hat \xi^{3},\hat x^{0}] = \frac {1}{\kappa}  \hat \xi^{3}    $ \\[4pt]
  &  $  [\hat \theta^{1},\hat x^{2}] = \frac {1}{\kappa}  \hat \xi^{3}\quad\,    [\hat \theta^{2},\hat x^{1}] = -\frac {1}{\kappa}  \hat \xi^{3}    $ & 0 &  $  [\hat \theta^{1},\hat x^{2}] = \frac {1}{\kappa}  \hat \xi^{3}\quad\,    [\hat \theta^{2},\hat x^{1}] = -\frac {1}{\kappa}  \hat \xi^{3}    $  \\[4pt]
$\mathcal{L}_{\kappa}$ &\redxmark & \redxmark & 0 \\[4pt]

\hline

\hline
\end{tabular}
 \end{center}
}
 \end{table}


\subsect{From coisotropic Poisson homogeneous spaces to noncommutative spaces}
\label{s33}
 
 We recall that coboundary Lie bialgebras $(\mf g, \delta)$ are the tangent counterpart of coboundary Poisson--Lie groups $(G,\Pi)$ \cite{ChariPressley1994}, where the Poisson structure $\Pi$ on $G$ is given by the so-called Sklyanin bracket
\begin{align}
\begin{split}
\label{eq:sklyanin}
&\{f_1,f_2\}=r^{ij}\left( X^L_i f_1\, X^L_j f_2 - X^R_i f_1 \, X^R_j f_2 \right),\qquad f_1,f_2 \in \mathcal C (G),
\end{split}
\end{align} 
such that    $X^L_i$ and $ X^R_i$ are   left- and right-invariant vector fields  defined by
\begin{align}
\label{eq:ivf}
X^L_i f(g)&=\frac{\dd}{\dd t}\biggr\rvert _{t=0} f\left(g\, {\rm e}^{t Y_i}\right),  \qquad  X^R_i f(g)=\frac{\dd}{\dd t}\biggr\rvert _{t=0} f\left({\rm e}^{t Y_i} g\right),
\end{align}
 where $f \in \mathcal C (G)$, $g \in G$ and $Y_i \in \mathfrak g$.   The quantization (as a Hopf algebra) of   
the Poisson--Lie group $(G,\Pi)$ is just the corresponding quantum group.

A Poisson homogeneous space $(G/H,\pi)$ for a  Poisson--Lie group $(G,\Pi)$  is the classical homogeneous space $G/H$ (like (\ref{ai})) endowed with a Poisson structure $\pi$ which is covariant under the action of the Poisson--Lie group $(G,\Pi)$. In the case of a coisotropic Poisson homogeneous space (so $\mathfrak h$ satisfies~\eqref{coisotropy}), the Poisson structure $\pi$ on $G/H$ is straightforwardly derived through canonical projection from the Poisson--Lie structure $\Pi$ on $G$. Noncommutative spaces  can  be finally obtained as    quantizations   of  coisotropic Poisson homogeneous spaces.

The steps of the procedure to construct all allowed $\kappa$-noncommutative spaces associated with  Poincar\'e   homogeneous spaces  (\ref{ai}), as shown in Table~\ref{table3} (\emph{i.e.}~ten cases), are summarized as follows:
\begin{itemize}

\item Consider a faithful representation $\rho$ of the Poincar\'e   algebra $\mf g$.

\item Obtain by exponentiation a generic element  of the Poincar\'e group $G$ with the appropriate order. This means that for the generic   left coset space $M^\ell=G/{H}$ (\ref{tp1}) the corresponding Poincar\'e group element $G_{M^\ell}$ must be constructed in the form
\be
G_{M^\ell}= \exp{\!\bigl( t^1 \rho(T_1) \bigr)}  \cdots\, \exp\!{ \bigl( t^\ell \rho(T_\ell) \bigr)} H ,
\label{group2}
\ee
where $T_1,  \dots, T_\ell$ are the translation generators on $M^\ell$ (\ref{tp22}) and $H$ is the $(10-\ell)$D isotropy subgroup.

\item Calculate the corresponding left- and right-invariant vector fields (\ref{eq:ivf})  from $G_{M^\ell}$ (\ref{group2}).

\item Compute the Poisson brackets among the local translation       coordinates $(t^1,\dots,t^\ell)$  by  applying the 
 Sklyanin bracket (\ref{eq:sklyanin}) from a classical $r$-matrix given in Table~\ref{table1}. The resulting expressions define the  coisotropic Poisson homogeneous space.

\item Finally, quantize the Poisson homogeneous space thus obtaining the  noncommutative space in terms of the quantum coordinates 
$(\hat t^1,\dots,\hat t^\ell)$.

\end{itemize}

As an instructive warming-up application of this methodology, in the next section we provide the explicit derivation of the three well-known $\kappa$-Minkowski spacetimes.


\subsect{The construction of $\kappa$-Minkowski spacetimes}
\label{s34}

Let us now apply the above approach to the   (3+1)D Minkowski spacetime  $\mathcal{M}  =G/H_\st$ (\ref{ai}).   We consider the faithful representation $\rho : \mathfrak g  \rightarrow \text{End}(\mathbb R ^5)$ for a generic  element $X\in \mathfrak g $   given, in the kinematical basis (\ref{aa}), by 
\begin{equation}
\rho(X)=   x^\mu \rho(P_\mu)  +  \xi^a \rho(K_a) +  \theta^a \rho(J_a) =
\left(\begin{array}{ccccc}
0&0&0&0&0\cr 
x^0 &0&\xi^1&\xi^2&\xi^3\cr 
x^1 &\xi^1&0&-\theta^3&\theta^2\cr 
x^2 &\xi^2&\theta^3&0&-\theta^1\cr 
x^3 &\xi^3&-\theta^2&\theta^1&0
\end{array}\right) \, ,
\label{ab}
\end{equation}
and the corresponding exponential map provides a 5D representation of the Poincar\'e group $G$. According to (\ref{group2}),  we construct   an element of the Poincar\'e group $G$ in the form
\begin{align}
\label{eq:Gm}
G_\mathcal{M}= \exp{\!\bigl(x^0 \rho(P_0)\bigr)} \exp{\!\bigl(x^1 \rho(P_1)\bigr)} \exp{\!\bigl(x^2 \rho(P_2)\bigr)} \exp{\!\bigl(x^3 \rho(P_3)\bigr)} \, H_\st,
\end{align}
where the Lorentz subgroup $ H_\st={\rm SO}(3,1)$ is parametrized by
\begin{align}
\label{eq:Lm}
  H_\st= \exp\bigl({\xi^1 \rho(K_1)}\bigr) \exp\bigl({\xi^2 \rho(K_2)}\bigr) \exp\bigl({\xi^3 \rho(K_3)}\bigr) \exp\bigl({\theta^1 \rho(J_1)}\bigr) \exp\bigl({\theta^2 \rho(J_2)} \bigr)\exp\bigl({\theta^3 \rho(J_3)}\bigr) .
\end{align}
From this, we compute the corresponding  left- and right-invariant vector fields  (\ref{eq:ivf}). The  Sklyanin bracket (\ref{eq:sklyanin})  for 
$ r_\timek$ and $ r_\spa$,  expressed in the kinematical basis given in Table~\ref{table1}, gives rise in this case  to linear Poisson brackets for the classical coordinates $x^\mu$. Therefore these Poisson brackets can be directly quantized, thus defining the  Lie-algebraic $\kappa$-Minkowskian spacetimes~\cite{BP2014extendedkappa,Maslanka1993,GH2021symmetry} with quantum coordinates  $\hat x^\mu$ which are shown in Table~\ref{table4}.

In the null-plane basis (\ref{ae}) we consider the following representation $\rho : \mathfrak g  \rightarrow \text{End}(\mathbb R ^5)$ for a generic  element $X\in \mathfrak g $~\cite{Rnullplane97}:
\bea
&&  \rho(X)=   x^+ \rho(P_+) + x^- \rho(P_-) + x^i  \rho(P_i)+ e^i  \rho(E_i)+ f^i  \rho(F_i)+\xi^3  \rho(K_3)+\phi^3  \rho(L_3) \nonumber\\[4pt]
&&\qquad\  \, =\left(\begin{array}{ccccc}
0&0&0&0&0\\[2pt]
\frac 12 x^+ + x^- &0&\frac12 e^1 +f^1 &\frac12 e^2 +f^2&\xi^3\\[2pt]
x^1 &\frac12 e^1 +f^1&0&\phi^3&-\frac12 e^1 +f^1\\[2pt]
x^2 &\frac12 e^2 +f^2&-\phi^3&0&-\frac12 e^2 +f^2\\[2pt]
\frac 12 x^+ - x^-&\xi^3&\frac12 e^1 -f^1&\frac12 e^2 -f^2&0
\end{array}\right) \, .
\label{af2}
\eea
And    the Poincar\'e group element is obtained  as
 \be
G_\mathcal{M}= \exp\!\left( {x^+ \rho(P_+)} \right) \exp\!\left({x^1 \rho(P_1)}  \right)\exp\!\left( {x^2 \rho(P_2)}  \right)  \exp\!\left( {x^- \rho(P_-)} \right) H_\st  ,
 \label{cb}
 \ee
where $H_\st$ is the Lorentz subgroup in the null-plane basis. Note that the relations between the spacetime coordinates   in the representation  (\ref{ab})  and those in (\ref{af2}) read
\be
x^0=\tfrac12   x^+ + x^- ,\qquad x^3= \tfrac12   x^+ - x^-,\qquad x^+= x^0+x^3,\qquad x^-=\tfrac12 (x^0-x^3),
 \label{cb2}
\ee
keeping $x^1$ and $x^2$.

Again,  if we now calculate the corresponding  invariant vector fields  (\ref{eq:ivf}) and next compute
 the Sklyanin bracket (\ref{eq:sklyanin}) for $ r_\nullplane$  written in the null-plane basis in Table~\ref{table1}, then we  obtain linear 
Poisson brackets for the classical coordinates $x^\alpha$  $(\alpha=\pm,1,2)$, whose trivial quantization gives rise  to the light-like or null-plane $\kappa$-Minkowskian spacetime~\cite{BP2014extendedkappa,Rnullplane97}  with quantum coordinates  $\hat x^\alpha$  given in Table~\ref{table4}.


\sect{Noncommutative spaces of geodesics}
\label{s4}

\subsect{Quantum time-like geodesics}
\label{s41}

This section contains the main results of this paper, namely the complete noncommutative spaces of time-  and space-like geodesics (corresponding to the second and third columns of Table \ref{table3}). For the sake of clarity, we firstly summarize (see~\cite{BGH2019worldlinesplb} for a detailed discussion)
 the construction of the homogeneous space of time-like geodesics, \emph{i.e.}~the geodesics followed by massive free particles on  Minkowski spacetime. In order to do that we use the second decomposition of the Poincar\'e Lie algebra given in \eqref{ai}, and we use a parametrization of the Poincar\'e Lie group (\ref{group2}) adapted to this coset space $\mathcal{W}_{\til} =  G / H_{\til}$ in the kinematical basis (\ref{aa}). In particular, we parametrize the Poincar\'e Lie group from the 5D matrix representation (\ref{ab}) as
\begin{align}
\begin{split}
\label{eq:Gtl}
&G_{\mathcal{W}_{\til} } = \exp\bigl({\eta^1 \rho(K_1)}\bigr) \exp\bigl({y^1 \rho(P_1)}\bigr) \exp\bigl({\eta^2 \rho(K_2)}\bigr) \exp\bigl({y^2 \rho(P_2)} \bigr)\exp\bigl({\eta^3 \rho(K_3)}\bigr) \exp\bigl({y^3 \rho(P_3)}\bigr) \,H_{\til} ,
\end{split}
\end{align}
where $H_{\til}$ is    the stabilizer  of the worldline corresponding to a massive particle at rest at the origin of $\mathcal M$, namely
\begin{align}
\begin{split}
\label{eq:Htl}
&H_{\til} = \exp\bigl( {\phi^1 \rho(J_1)}\bigr) \exp\bigl( {\phi^2 \rho(J_2)}\bigr) \exp\bigl( {\phi^3 \rho(J_3)}\bigr) \exp\bigl( {y^0 \rho(P_0)} \bigr).
\end{split}
\end{align}
In this way $( y^a, \eta^a)$ provide a set of coordinates on the 6D space  $\mathcal{W}_{\til}$. At the level of the Lie group, the time-like geodesic parametrization \eqref{eq:Gtl} and the spacetime parametrization \eqref{eq:Gm} are related by 
\be
x^\mu  = x^\mu (y^\mu, \eta^a) , \qquad \xi^a = \eta^a ,\qquad  \theta^a  =  \phi^a ,
\label{xf}
\ee
where
\begin{align}
\begin{split}
\label{eq:falpha}
&x^0 = y^1 \sinh \eta^1 + \cosh \eta^1 \left( y^2 \sinh \eta^2 + \cosh \eta^2 (y^0 \cosh \eta^3 + y^3 \sinh \eta^3) \right) ,\\
&x^1  = y^1 \cosh \eta^1 + \sinh \eta^1 \left( y^2 \sinh \eta^2 + \cosh \eta^2 (y^0 \cosh \eta^3 + y^3 \sinh \eta^3) \right) ,\\
&x^2  = y^2 \cosh \eta^2 + \sinh \eta^2 (y^0 \cosh \eta^3 + y^3 \sinh \eta^3), \\
&x^3  = y^0 \sinh \eta^3 + y^3 \cosh \eta^3 \, ,
\end{split}
\end{align}
so involving the six coordinates on the space of time-like geodesics plus the `extra' coordinate $y^0$.

In the following, we present the three noncommutative spaces of time-like geodesics defined by the $\kappa$-Poincar\'e family of $r$-matrices given in Table \ref{table1} by applying the very same procedure to the one described in     section~\ref{s34} to construct the three  $\kappa$-Minkowski spacetimes. Thus we   compute the left- and right-invariant vector fields (\ref{eq:ivf}) from $G_{\mathcal{W}_{\til} } $ (\ref{eq:Gtl}) and  
obtain the   Poisson--Lie structure on the Poincar\'e group associated to a given Lie bialgebra by means of the Sklyanin bracket (\ref{eq:sklyanin}). Then we project the Poisson--Lie structure to the coset space $\mathcal{W}_{\til}$ in order to get the coisotropic Poisson structure of time-like worldlines and, finally, we quantize the Poisson homogeneous space to obtain the algebra of `quantum geodesic' observables $\mathcal{W}_{\til,\kappa}$.


\subsubsect{From the time-like $\kappa$-deformation}
\label{s411}

Here we sketch the essential results given in~\cite{BGH2019worldlinesplb}.
If we consider the well-known time-like $\kappa$-Poincar\'e $r$-matrix~\cite{Maslanka1993}  $r_\timek$ and we follow the above-mentioned procedure, then  by projecting the Sklyanin bracket to the homogeneous space coordinates we get a coisotropic structure for the classical space of time-like geodesics which can be straightforwardly quantized, since no ordering problems appear.  In this way, the quantum space  of time-like geodesics $\mathcal{W}_{\til,\kappa}$ reads:
\begin{equation}
\begin{split}
[\hat y^1, \hat y^2]_t &= \frac{1}{\kappa} \left( \hat y^2 \sinh \hat \eta^1-\frac{\hat y^1 \tanh \hat \eta^2}{\cosh \hat \eta^3}\right) ,\\
[\hat y^1,\hat y^3]_t &= \frac{1}{\kappa} \big(\hat y^3 \sinh  \hat \eta^1 -\hat y^1 \tanh  \hat \eta^3 \big), \\
[\hat y^2, \hat y^3]_t &= \frac{1}{\kappa} \big(\hat y^3 \cosh  \hat \eta^1  \sinh  \hat \eta^2 -\hat y^2 \tanh  \hat \eta^3 \big), \\
[\hat y^1, \hat \eta^1]_t &= \frac{1}{\kappa} \,\frac{ \bigl(\cosh  \hat \eta^1  \cosh  \hat \eta^2  \cosh  \hat \eta^3 -1\bigr)}{\cosh  \hat \eta^2  \cosh  \hat \eta^3 }, \\
[\hat y^2, \hat \eta^2]_t &= \frac{1}{\kappa}\, \frac{ \bigl(\cosh  \hat \eta^1  \cosh  \hat \eta^2  \cosh  \hat \eta^3 -1\bigr)}{\cosh  \hat \eta^3 }, \\
[\hat y^3, \hat \eta^3]_t &= \frac{1}{\kappa}\, \bigl(\cosh  \hat \eta^1  \cosh  \hat \eta^2  \cosh  \hat \eta^3 -1\bigr) \, ,
\label{eq:comm_t_rt}
\end{split}
\end{equation}
together with
\be
[\hat \eta^a, \hat\eta^b]_t = 0,\quad  \forall a,b,\qquad [\hat y^a,\hat \eta^b]_t = 0,\quad    a \neq b. 
\label{xz1}
\ee
It is remarkable that the coisotropic Poisson structure given by the Poisson brackets identical to~\eqref{eq:comm_t_rt} is found to be symplectic almost everywhere, \emph{i.e.}~it is symplectic in the 6D smooth submanifold where $(\eta^1,\eta^2,\eta^3) \neq (0,0,0)$. In fact, the new coordinates
\begin{equation}
\begin{split}
 q^1_t &= \frac{ y^1 \cosh   \eta^2  \cosh   \eta^3 }{\cosh   \eta^1  \cosh   \eta^2  \cosh   \eta^3 -1} , \\
 q^2_t &= \frac{ y^2 \cosh   \eta^3 }{\cosh   \eta^1  \cosh   \eta^2  \cosh   \eta^3 -1} , \\
 q^3_t &= \frac{ y^3}{\cosh   \eta^1  \cosh   \eta^2  \cosh   \eta^3 -1} , \\
 p^a &=  \eta^a ,
\label{eq:darboux_t_rt}
\end{split}
\end{equation}
can be considered as the Darboux coordinates on such submanifold, since
their Poisson brackets read
\begin{align}
\label{eq:pois_canonical}
\{q_t^a,q_t^b\}= \{p^a,p^b\}= 0, \qquad   \{q_t^a,p^b\}= \frac 1 {\kappa} \, \delta_{ab} \, .
\end{align}
Therefore, the quantum counterpart of~\eqref{eq:darboux_t_rt} leads to the usual noncommutative phase space algebra of quantum mechanics where the deformation parameter $\kappa^{-1}$ replaces the Planck constant $\hbar$.


\subsubsect{From the space-like $\kappa$-deformation}
\label{s412}

Similarly to the previous case, we can repeat the same procedure to obtain the Poisson homogenous structure on the space of time-like geodesics associated to the second $r$-matrix  $r_\spa$ from Table \ref{table1}. The Poisson structure so obtained is evidently different from the one obtained before, but it has clear formal analogies: it admits a trivial quantization (no ordering problems arise again), and it is almost everywhere symplectic. In particular, the non-vanishing brackets (already quantized) defining $\mathcal{W}_{\til,\kappa}$  from $ r_\spa$  read
\begin{equation}
\begin{split}
[\hat y^1,\hat y^2]_s &= - \frac{1}{\kappa}\, \hat y^1 \tanh  \hat \eta^2  \tanh  \hat \eta^3,  \\
[\hat y^1, \hat y^3]_s &= \frac{1}{\kappa}\, \frac{\hat y^1}{\cosh  \hat \eta^3 } ,\\
[\hat y^2, \hat y^3]_s &= \frac{1}{\kappa} \,\frac{\hat y^2}{\cosh  \hat \eta^3 }, \\
[\hat y^1, \hat \eta^1]_s &= -\frac{1}{\kappa}\, \frac{\tanh  \hat \eta^3 }{\cosh  \hat \eta^2 } ,\\
[\hat y^2, \hat \eta^2]_s &= -\frac{1}{\kappa} \tanh  \hat \eta^3  ,\\
[\hat y^3, \hat \eta^3]_s &= -\frac{1}{\kappa} \sinh  \hat \eta^3 .
\label{eq:comm_t_rs}
\end{split}
\end{equation}
The classical Darboux coordinates (which now are only defined in the submanifold $\eta^3 \neq 0$) take the form
\begin{equation}
\begin{split}
 q^1_s &= -\frac{ y^1 \cosh   \eta^2 \cosh   \eta^3}{\sinh   \eta^3} , \\
 q^2_s &= -\frac{ y^2 \cosh   \eta^3}{\sinh   \eta^3} , \\
 q^3_s &= -\frac{ y^3}{\sinh   \eta^3} , \\
 p^a &=  \eta^a ,
\label{eq:darboux_t_rs}
\end{split}
\end{equation}
 and fulfil again the canonical Poisson brackets (\ref{eq:pois_canonical}).


\subsubsect{From the light-like $\kappa$-deformation}
\label{s413}

In section \ref{s2} it was already discussed that the light-like $\kappa$-Poincar\'e $r$-matrix  $ r_\nullplane$ in Table~\ref{table1} is   obtained as the sum of the time-like and space-like ones \eqref{bbaa2}. Thus, its associated Lie bialgebra  and Poisson--Lie structure  are given by the linear superposition of the two structures that we have just found. Again, it is immediate to realize that no ordering problems appear and the full quantization can be performed directly, giving rise to a quantum space $\mathcal{W}_{\til,\kappa}$ defined by
\begin{equation}
\begin{split}
[\hat y^1,\hat y^2]_l &= \frac{1}{\kappa} \left(\hat y^2 \sinh  \hat \eta^1 -\frac{\hat y^1  \tanh \hat \eta^2  \bigl( \sinh  \hat \eta^3 +1\bigr)}{\cosh  \hat \eta^3 }\right) ,\\
[\hat y^1,\hat y^3]_l &= \frac{1}{\kappa} \left( \hat y^3 \sinh  \hat \eta^1 -\frac{\hat y^1 \bigl( \sinh  \hat \eta^3 -1\bigr)}{\cosh  \hat \eta^3 }\right) ,\\
[\hat y^2, \hat y^3]_l &= \frac{1}{\kappa} \left(\hat y^3 \cosh  \hat \eta^1  \sinh  \hat \eta^2 -\frac{\hat y^2 (\sinh  \hat \eta^3 -1)}{\cosh  \hat \eta^3 }\right),\\
[\hat y^1, \hat \eta^1]_l &= \frac{1}{\kappa} \left( \frac{ \cosh  \hat \eta^1  \cosh  \hat \eta^2  \cosh  \hat \eta^3-\sinh  \hat \eta^3 -1}{\cosh  \hat \eta^2  \cosh  \hat \eta^3 }  \right), \\
[\hat y^2, \hat \eta^2]_l &= \frac{1}{\kappa} \left( \frac{\cosh  \hat \eta^1  \cosh  \hat \eta^2  \cosh  \hat \eta^3 - \sinh  \hat \eta^3 -1}{\cosh  \hat \eta^3 } \right) ,\\
[\hat y^3, \hat \eta^3]_l &= \frac{1}{\kappa} \left( \cosh  \hat \eta^1  \cosh  \hat \eta^2  \cosh  \hat \eta^3- \sinh  \hat \eta^3   -1   \right) . 
\label{eq:comm_t_rl2}
\end{split}
\end{equation}
The Darboux coordinates in this case are found to be
\begin{equation}
\begin{split}
 q^1_l &= \frac{ y^1 \cosh   \eta^2  \cosh   \eta^3 }{\cosh   \eta^1  \cosh   \eta^2  \cosh   \eta^3 -\sinh  \hat \eta^3 -1} ,\\
 q^2_l &= \frac{ y^2 \cosh   \eta^3 }{\cosh   \eta^1  \cosh   \eta^2  \cosh   \eta^3 -\sinh   \eta^3 -1}, \\
 q^3_l &= \frac{ y^3}{\cosh   \eta^1  \cosh   \eta^2  \cosh   \eta^3 -\sinh   \eta^3 -1} ,\\
 p^a &=  \eta^a ,
\label{eq:darboux_t_rl}
\end{split}
\end{equation}
 satisfying the canonical Poisson brackets (\ref{eq:pois_canonical}).

We stress that under linearization, that is, by considering the first-order in the quantum coordinates  $(\hat y^a,\hat \eta^a)$    of the three   quantum spaces $\mathcal{W}_{\til,\kappa}$ (\ref{eq:comm_t_rt}), (\ref{eq:comm_t_rs}) and ({\ref{eq:comm_t_rl2}), we recover the first-order  noncommutative spaces given in the second row in Table~\ref{table4}, provided that, only at this first-order, the quantum coordinates are given by $\hat x^\mu\equiv \hat y^\mu$ and $\hat \xi^a\equiv \hat \eta^a$ (see (\ref{xf}) and (\ref{eq:falpha})). In this respect, also observe  that if we compute the above Poisson brackets from the three $r$-matrices for the classical  spacetime coordinates  $x^\mu$ given by  (\ref{eq:falpha}) we recover the three   linear  $\kappa$-Minkowski spacetimes shown in the first row of Table~\ref{table4}. This makes evident that the full quantum spaces (\ref{eq:comm_t_rt}), (\ref{eq:comm_t_rs}) and ({\ref{eq:comm_t_rl2}) are defined by strongly non-linear relations that can only be  envisaged once the complete classical coisotropic Poisson structures are computed.

It is also worth noticing that the three sets of Darboux coordinates  $q_t^a$ \eqref{eq:darboux_t_rt},  $q_s^a$ \eqref{eq:darboux_t_rs} and  $q_l^a$ \eqref{eq:darboux_t_rl}    are connected through the simple relation
\begin{equation}
\frac{1}{ q^a_l} = \frac{1}{ q^a_t} + \frac{1}{ q^a_s} ,
\label{eq:relation_darboux_tl}
\end{equation}
while Darboux momenta coincide for the three cases $ p^a=  \eta^a$. Observe also that in none of the three cases the definition of the Darboux coordinates $q^a$ can be linearized in terms of the local coordinates on the given space, which is consistent with the fact that the first-order noncommutative spaces given in Table~\ref{table4} are  not isomorphic as Lie algebras to the three Heisenberg--Weyl algebras quantizing~\eqref{eq:pois_canonical}.


\subsect{Quantum space-like geodesics}
\label{s42}

Let us now describe the noncommutative spaces of  geodesics corresponding to the three families of space-like $\kappa$-Poincar\'e deformations from Table \ref{table1}. Similarly to the previous section, out first task is to parametrize the Poincar\'e Lie group in such a manner that we obtain a set of coordinates that descend to the appropriate quotient space $\mathcal{W}_{\sil} =  G / H_{\sil}$  \eqref{ai}. Such a parametrization is given by
\begin{align}
\begin{split}
\label{eq:Gsl}
&G_{\mathcal{W}_{\sil}} = \exp\bigl({\pi^1 \rho(J_2)}\bigr)  \exp\bigl({u^1 \rho(P_1)}\bigr) \exp\bigl({\pi^2 \rho(J_1)}\bigr)\exp\bigl({u^2 \rho(P_2)} \bigr)\exp\bigl({\pi^0 \rho(K_3)}\bigr) \exp\bigl({u^0 \rho(P_0)}\bigr) \, H_{\sil} ,
\end{split}
\end{align}
where now the isotropy subgroup $H_{\sil}$ of a space-like line takes the form
\begin{align}
\begin{split}
\label{eq:Hsl}
&H_{\sil} = \exp\bigl({v^2 \rho(K_2)} \bigr)\exp\bigl({v^1 \rho(K_1)}\bigr) \exp\bigl({v^3 \rho(J_3)}\bigr) \exp\bigl({u^3 \rho(P_3)} \bigr) .
\end{split}
\end{align}
Therefore $(u^0, u^1,  u^2, \pi^0, \pi^1,\pi^2)$  define a set of coordinates on the 6D space $\mathcal{W}_{\sil}$ (see (\ref{group2})). It is useful to observe that, at the level of the Lie group, the    four `spacetime coordinates' $x^\mu$ in \eqref{eq:Gm} can be expressed in terms of the seven `space-like geodesic coordinates' $(u^\mu, \pi^0, \pi^1,\pi^2)$ in (\ref{eq:Gsl})  as
\begin{align}
\begin{split}
\label{eq:falpha2}
&x^0  = u^0 \cosh \pi^0+u^3 \sinh \pi^0 ,\\
&x^1  =u^1 \cos \pi^1+u^2 \sin \pi^1 \sin \pi^2+ \sin \pi^1 \cos \pi^2 \bigl(u^0 \sinh \pi^0+u^3 \cosh \pi^0 \bigr) ,\\
&x^2  = u^2 \cos \pi^2-\sin \pi^2 \bigr(u^0 \sinh \pi^0+u^3 \cosh \pi^0 \bigl), \\
&x^3  = \cos \pi^1 \cos \pi^2 \bigr(u^0 \sinh \pi^0+u^3 \cosh \pi^0 \bigr)-u^1 \sin \pi^1+u^2 \cos \pi^1 \sin \pi^2 \, .
\end{split}
\end{align}
Note that, similarly to the time-like case, these relations do not only involve the six coordinates on the space of space-like geodesics $\mathcal{W}_{\sil}$, but also the `extra'  one $u^3$; in fact the linear approximation of (\ref{eq:falpha2}) gives that $x^\mu\equiv u^\mu$. The remaining translation coordinates  $(\pi^0, \pi^1,\pi^2)$     in (\ref{eq:Gsl})  correspond, at this first-order, to $(\xi^3,\theta^2,\theta^1)$ in \eqref{eq:Gm}.

In the rest of this section, we present the explicit noncommutative algebra of  quantum space-like geodesics, obtained as the quantization of the three Poisson homogeneous structures on   $\mathcal{W}_{\sil}$ which, in turn, is constructed by using the canonical projection $G_{\mathcal{W}_{\sil}}  \to \mathcal{W}_{\sil}$ and the fact that all these Poisson structures are coisotropic with respect to this quotient. 

The three noncommutative algebras of operators share a number of characteristics, which make them formally similar to the structures presented in the previous section. Remarkably, all of them share the same subset of vanishing commutators 
\be
[\hat \pi^m, \hat \pi^n]_k = 0,\quad m,n=0,1,2,\qquad [\hat u^m, \hat \pi^n]_k = 0,\quad m\ne n ,
\ee
where the label $k\in \{ t,s,l \}$ refers to the time-, space, and light-like $r$-matrix considered as in the previous section. These relations will be essential in order to avoid any ordering ambiguity under quantization.

 
\subsubsect{From the time-like $\kappa$-deformation}
\label{s421}

We start from the   $r$-matrix $ r_\timek$  of Table \ref{table1}  and after quantization of its coisotropic Poisson homogeneous structure we obtain that the quantum space of geodesics $\mathcal{W}_{\sil,\kappa}$  is defined by the following non-vanishing commutators
\begin{equation}
\begin{split}
[\hat u^0, \hat u^1]_t &= -\frac{1}{\kappa} \, \frac{\hat u^1}{\cosh \hat \pi^0} ,\\
[\hat u^0, \hat u^2]_t &= - \frac{1}{\kappa} \,  \frac{\hat u^2}{\cosh \hat \pi^0} ,\\
[\hat u^1, \hat u^2]_t &= \frac{1}{\kappa} \,  \hat u^1 \tanh \hat \pi^0 \tan \hat \pi^2, \\
[\hat u^0, \hat \pi^0]_t &= - \frac{1}{\kappa}   \sinh \hat \pi^0 ,\\
[\hat u^1, \hat \pi^1]_t &= \frac{1}{\kappa}  \, \frac{\tanh \hat \pi^0}{\cos \hat \pi^2} ,\\
[\hat u^2, \hat \pi^2]_t &= - \frac{1}{\kappa}   \tanh \hat \pi^0. 
\label{eq:comm_s_rt}
\end{split}
\end{equation}
The corresponding Poisson homogeneous structure can be expressed in canonical form on the 6D submanifold $\pi^0 \neq 0$ by means of the following change to Darboux-type coordinates
\begin{equation}
\begin{split}
 q^0_t &= -\frac{ u^0}{\sinh  \pi^0}, \\
 q^1_t &= \frac{ u^1 \cosh \pi^0\cos  \pi^2}{ \sinh  \pi^0}, \\
 q^2_t &= -\frac{ u^2 \cosh  \pi^0}{\sinh  \pi^0} ,\\
 p^m &=  \pi^m,
\label{eq:darboux_s_rt}
\end{split}
\end{equation}
such that a $\kappa$-canonical Poisson structure is obtained:
\begin{align}
\label{eq:pois_canonical2}
\{q_t^m,q_t^n\}= \{p^m,p^n\}= 0, \qquad   \{q_t^m,p^n\}= \frac 1 {\kappa} \, \delta_{mn}  ,\quad \quad m,n=0,1,2 .
\end{align}


\subsubsect{From the space-like $\kappa$-deformation}
\label{s422}

From the $\kappa$-Poincar\'e $r$-matrix $r_\spa$   we obtain, after quantization, that the noncommutative algebra of space-like geodesics $\mathcal{W}_{\sil,\kappa}$  is defined by the non-vanishing brackets
\begin{equation}
\begin{split}
[\hat u^0, \hat u^1]_s &= - \frac{1}{\kappa} \bigr(\hat u^0 \sin \hat \pi^1-\hat u^1 \tanh \hat \pi^0 \bigl), \\
[\hat u^0, \hat u^2]_s &= \frac{1}{\kappa} \bigr(\hat u^0 \cos \hat \pi^1 \sin \hat \pi^2+\hat u^2 \tanh \hat \pi^0 \bigl) ,\\
[\hat u^1, \hat u^2]_s &= \frac{1}{\kappa} \left(\frac{\hat u^1 \tan \hat \pi^2}{\cosh \hat \pi^0}+\hat u^2 \sin \hat \pi^1\right) ,\\
[\hat u^0, \hat \pi^0]_s &= \frac{1}{\kappa} \bigl(\cosh \hat \pi^0\cos \hat \pi^1 \cos \hat \pi^2-1 \bigr), \\
[\hat u^1, \hat \pi^1]_s &= - \frac{1}{\kappa} \,\frac{\bigl(\cosh \hat \pi^0 \cos \hat \pi^1 \cos \hat \pi^2-1 \bigr)}{\cosh \hat \pi^0 \cos \hat \pi^2} ,\\
[\hat u^2, \hat \pi^2]_s &= \frac{1}{\kappa}\, \frac{\bigl(\cosh \hat \pi^0 \cos \hat \pi^1\cos \hat \pi^2-1\bigr)}{\cosh \hat \pi^0} . 
\label{eq:comm_s_rs}
\end{split}
\end{equation}
In this case the Poisson homogeneous structure is non-degenerate on the 6D submanifold $(\pi^0,\pi^1,\pi^2) \neq (0,0,0)$, and the Darboux-type coordinates are given by
\begin{equation}
\begin{split}
 q^0_s &= \frac{ u^0}{\cosh  \pi^0 \cos \pi^1 \cos  \pi^2-1} ,\\
 q^1_s &= -\frac{ u^1 \cosh  \pi^0 \cos  \pi^2}{\cosh  \pi^0 \cos  \pi^1 \cos  \pi^2-1} ,\\
 q^2_s &= \frac{ u^2 \cosh \pi^0}{\cosh \pi^0 \cos  \pi^1 \cos  \pi^2-1} ,\\
 p^m &=  \pi^m ,
\label{eq:darboux_s_rs}
\end{split}
\end{equation}
verifying again the canonical Poisson brackets (\ref{eq:pois_canonical2}).


\subsubsect{From the light-like $\kappa$-deformation}
\label{s423}

Finally, for the   $r$-matrix  $r_\nullplane$ we obtain a Lie bialgebra and therefore, a Poisson homogeneous structure, that is the linear superposition of the ones associated with the time-  and space-like  deformations that we have just presented. Upon quantization, the non-vanishing brackets defining the quantum space $\mathcal{W}_{\sil,\kappa}$  read
\begin{equation}
\begin{split}
[\hat u^0, \hat u^1]_l &= - \frac{1}{\kappa} \left(\hat u^0 \sin \hat \pi^1-\frac{\hat u^1 \bigl(\sinh \hat \pi^0-1 \bigr)}{\cosh \hat \pi^0}\right) ,\\
[\hat u^0, \hat u^2]_l &= \frac{1}{\kappa} \left(\hat u^0 \cos \hat \pi^1 \sin \hat \pi^2+\frac{\hat u^2 \bigl(\sinh \hat \pi^0-1\bigr)}{\cosh \hat \pi^0}\right) ,\\
[\hat u^1, \hat u^2]_l &= \frac{1}{\kappa} \left(\frac{\hat u^1 \bigl(\sinh \hat \pi^0+1\bigr) \tan \hat \pi^2}{\cosh \hat \pi^0}+\hat u^2 \sin \hat \pi^1\right) ,\\
[\hat u^0, \hat \pi^0]_l &= \frac{1}{\kappa} \bigl(\cosh \hat \pi^0 \cos \hat \pi^1 \cos \hat \pi^2-\sinh \hat \pi^0-1 \bigr) ,\\
[\hat u^1, \hat \pi^1]_l &= - \frac{1}{\kappa} \,\frac{\bigl( \cosh \hat \pi^0 \cos \hat \pi^1 \cos \hat \pi^2-\sinh \hat \pi^0-1\bigr) }{\cosh \hat \pi^0 \cos \hat \pi^2} ,\\
[\hat u^2, \hat \pi^2]_l &= \frac{1}{\kappa} \,\frac{\bigl( \cosh \hat \pi^0 \cos \hat \pi^1 \cos \hat \pi^2-\sinh \hat \pi^0-1\bigr)}{\cosh \hat \pi^0} . 
\label{eq:comm_s_rl}
\end{split}
\end{equation}
Similarly to the previous case, this Poisson homogenous structure is symplectic when restricted to the 6D submanifold $(\pi^0,\pi^1,\pi^2) \neq (0,0,0)$, with Darboux coordinates given by
\begin{equation}
\begin{split}
 q^0_l &= \frac{ u^0}{\cosh  \pi^0\cos  \pi^1 \cos  \pi^2-\sinh  \pi^0-1} ,\\
 q^1_l &= -\frac{ u^1 \cosh  \pi^0 \cos \pi^2}{\cosh  \pi^0 \cos \pi^1\cos  \pi^2-\sinh \pi^0-1} ,\\
 q^2_l &= \frac{ u^2 \cosh  \pi^0}{\cosh \pi^0 \cos \pi^1 \cos  \pi^2-\sinh  \pi^0-1} ,\\
 p^m &=  \pi^m ,
\label{eq:darboux_s_rl}
\end{split}
\end{equation}
fulfilling once more (\ref{eq:pois_canonical2}).

If we consider  the first-order in the quantum coordinates $(\hat u^m, \hat \pi^m)$  $(m=0,1,2)$ in  the three   quantum spaces $\mathcal{W}_{\sil,\kappa}$ (\ref{eq:comm_s_rt}), (\ref{eq:comm_s_rs}) and ({\ref{eq:comm_s_rl}), we recover, as expected, the first-order  noncommutative spaces given in the third row in Table~\ref{table4}, by taking into account that in such linear approximation $(\hat x^m, \hat \xi^3, \hat \theta^2,\hat \theta^1)$ coincide with 
 $(\hat u^m, \hat \pi^0, \hat \pi^1,  \hat \pi^2)$, respectively. In addition,  when the above Poisson structures from the three $r$-matrices  are calculated for the classical  spacetime coordinates  $x^\mu$ when expressed as the functions (\ref{eq:falpha2}), the Poisson version of the three   linear  $\kappa$-Minkowski spacetimes shown in the first row of Table~\ref{table4} is recovered, thus showing the complete self-consistency of this approach.

It is interesting to note that a similar relation to \eqref{eq:relation_darboux_tl} is satisfied among the Darboux coordinates \eqref{eq:darboux_s_rt}, \eqref{eq:darboux_s_rs} and \eqref{eq:darboux_s_rl} for the three different $\kappa$-Poincar\'e Poisson homogeneous structures on the space of space-like geodesics, namely 
\begin{equation}
\frac{1}{q^m_l} = \frac{1}{q^m_t} + \frac{1}{q^m_s} ,\qquad m=0,1,2.
\label{eq:relation_darboux_sl}
\end{equation}


\subsect{Quantum light-like geodesics}
\label{s43}

For the sake of completeness, we summarize here the defining relations for the 5D  noncommutative space of light-like geodesics $\mathcal{L}_\kappa$  defined by the classical $r$-matrix $ r_\nullplane$ in Table~\ref{table1}, which has recently been  constructed and analysed in~\cite{BGH2022light}. We insist on the fact that this is the only light-like space of geodesics that can be constructed, since $ r_\timek$ and $ r_\spa$ do not provide a coisotropic Poisson homogeneous structure amenable to be quantized.

In this case the Poincar\'e group representation is constructed in the null-plane basis (\ref{af2}) and reads
\be
G_\mathcal{L}= \exp\left( {y^- \rho(P_-)} \right)\exp\left({f^1 \rho(F_1)}  \right)\exp\left( {f^2 \rho(F_2)} \right)  \exp\left({y^1 \rho(P_1)}  \right)\exp\left( {y^2 \rho(P_2)}  \right)  H_\lil  ,
 \label{ce}
 \ee
where  $H_\lil$ is the isotropy subgroup of a light-like line, which in this case reads (see~\cite{BGH2022light} for details)
\be
 H_\lil= \exp\left({y^+ \rho(P_+)}  \right)\exp\left({e^1 \rho(E_1)}  \right)\exp\left( {e^2 \rho(E_2)} \right)  \exp\left({\xi^3 \rho(K_3)} \right)  \exp\left({\phi^3 \rho(L_3)} \right)   .
\label{cf}
\ee 
Then  the generators $\{ P_-,P_i, F_i\}$ $(i=1,2)$ give rise to translations on the 5D   homogeneous space $\mathcal{L}  =G/H_\lil $ (\ref{ai}) which is thus parametrized in terms of the coordinates $(y^-,y^i, f^i)$. Then the coisotropic Poisson homogeneous structure can be straightforwardly computed and quantized, giving rise to the following commutation rules that define the quantum space of $\kappa$-Poincar\'e light-like geodesics $\mathcal{L}_\kappa$:
  \be
\begin{array}{ll}
\displaystyle{ \bigl[\hat y^-,\hat  y^i \bigr]=-\frac 2\kappa \, \hat  f^i \hat  y^- , }&\quad \displaystyle{ \bigl[\hat y^1,\hat y^2\bigr]= \frac 2\kappa\bigl( \hat f^1 \hat y^2 - \hat  f^2\hat y^1\bigr) , }\\[6pt] 
\displaystyle{ \bigl[\hat y^i, \hat f^j \bigr]= \frac 1\kappa \,\delta_{ij}   \bigl( (\hat f^1)^2+  (\hat f^2)^2\bigr) , }&\quad  \bigl[\hat y^-,\hat f^i \bigr]=  \bigl[\hat f^1,\hat f^2 \bigr]=0,\qquad i,j=1,2.
 \end{array}
\label{ch}
\ee
Note that all these commutators are given by homogeneous quadratic  expressions, and therefore their linearization is zero, 
as Table~\ref{table4} predicts. In fact, this is an illustrative instance of the fact that, in some cases, the linear approximation to quantum spaces given by the dual of the cocommutator map provides a very limited amount of information. Finally it is worth mentioning that, as it was shown in~\cite{BGH2022light}, on a given irreducible representation fixed by a value of the Casimir operator $\hat C$ of the algebra~\eqref{ch}, a nonlinear change of basis transforms~\eqref{ch} into
\be
\bigl[ \hat q^i ,  \hat p^j\bigr]=\frac{1}{\kappa}\,\delta_{ij} ,\qquad
\bigl[ \hat y^- ,  \hat p^i\bigr]=0\, ,
\qquad
\bigl[ \hat y^- ,  \hat q^i\bigr]= - \frac{2\,\hat C}{\kappa}\, \hat p^i\, ,
\qquad i=1,2  .
\ee
Therefore,  the (5D)  algebra $\mathcal{L}_\kappa$ can be interpreted as  a non-central extension, generated by $\hat y^-$, of a direct sum of two Heisenberg--Weyl algebras $(\hat q^i,\hat p^i)$, while all the (6D) noncommutative spaces of time- and space-like lines previously constructed can be written as the direct sum of  three Heisenberg--Weyl algebras.


\sect{Concluding remarks}
\label{s5}

This paper has been focused on showing that the construction of noncommutative spaces of geodesics with quantum group symmetry can be explicitly performed by quantizing the coisotropic Poisson homogeneous spaces of geodesics that can be associated to each quantum deformation via its underlying classical $r$-matrix. Here we have explicitly considered the quantum geodesics on Minkowski spacetime arising from the $\kappa$-Poincar\'e quantum group, but the method is fully applicable to any other quantum deformation of the Poincar\'e group (for instance, to those ones endowed with a quantum Lorentz subgroup~\cite{Ballesteros:2021inq}), as well as to the spaces of geodesics and quantum deformations of other kinematical groups.

In this setting, the coisotropy condition for the Lie bialgebra associated to a given quantum deformation with respect to the isotropy subgroup of the chosen space of geodesics turns out to be the key property that allows the construction to be performed. When this condition is analyzed for the three possible $\kappa$-deformations of the Poincar\'e group, the first remarkable finding is the fact that the light-like or null-plane quantum Poincar\'e deformation is the only one that enables the construction of the three (time-, space- and light-like) quantum spaces of geodesics, since the time-like and space-like $\kappa$-deformations do not fulfil the coisotropy condition for the space of geodesics on the light cone.

Therefore, we have explicitly presented the seven possible $\kappa$-Poincar\'e coisotropic Poisson homogeneous spaces of geodesics and their quantization, by emphasizing among them the five which are new ones,  and by sketching under a common framework the other two. From a mathematical viewpoint, we stress that this construction can be performed provided that for each space of geodesics a careful choice of the parametrization of the Poincar\'e group leading to suitable coordinates on the homogeneous space is found, and for the case of the light-like deformation the so called null-plane basis of the Poincar\'e algebra is the natural one to be used. In addition, we have also emphasized the fact that the seven quantum spaces of geodesics are all of them defined by nonlinear algebras which are non-trivial higher order generalizations of the first-order noncommutative spaces that can be directly obtained from the dual Lie bialgebra structure. 

The results here presented open a number of questions which are worth to be faced in the near future. For instance, the finding that Darboux-type coordinates can be given within the spaces of geodesics paves the way to the analysis, following~\cite{LMMP2018localization,BGGM2021fuzzy}, of the fuzzy properties of all the types of quantum worldlines here introduced as well as the fuzziness of the events that can be defined from them. In particular, this would be specially interesting for the three quantum spaces of worldlines that can be obtained from the null-plane deformation, since in this case the quantum space of light-like geodesics should be obtained as a common `boundary' of the time- and space-like spaces. Moreover, the construction of the spaces of noncommutative (anti-)de Sitter geodesics which are covariant under the   already known  $\kappa$-(A)dS quantum groups (see~\cite{BHOS1994global,BHMN2017kappa3+1,BGGH2018CMS31,BGH2019kappaAdS3+1}) could be achieved by following the very same approach here presented, and in this way the role of a nonvanishing cosmological constant could be analysed in this novel noncommutative geometric setting. Work on all these lines is in progress.


\section*{Acknowledgements}

\phantomsection
\addcontentsline{toc}{section}{Acknowledgements}

This work has been partially supported by Agencia Estatal de Investigaci\'on (Spain)  under grant  PID2019-106802GB-I00/AEI/10.13039/501100011033, by the Regional Government of Castilla y Le\'on (Junta de Castilla y Le\'on, Spain) and by the Spanish Ministry of Science and Innovation MICIN and the European Union NextGenerationEU/PRTR. The authors would like to acknowledge the contribution of the European Cooperation in Science and Technology COST Action CA18108.


\small


\end{document}